\documentclass[twocolumn,superscriptaddress,showpacs]{revtex4}
\usepackage{graphicx}
\usepackage{dcolumn,xcolor,ulem}
\usepackage{amsmath} 
\usepackage{amssymb}
\usepackage{bm}
\usepackage[utf8]{inputenc} % utf8 est normalement valable partout (utf8x si qq charac merdent), tu peux aussi utiliser, latin1 sous unix, ansinew sous windows, applemac sous mac os
\usepackage[T1]{fontenc}
\usepackage{mathrsfs}

\begin{document}

\widowpenalty=1000
\clubpenalty=1000

%\preprint{APS/123-QED}

\title{Twisting and buckling: a new undulation mechanism for artificial swimmers}

\author{Ghani Oukhaled}
\affiliation{Laboratoire Mati\`ere et Syst\`emes Complexes, UMR 7057\\ CNRS $\&$ Universit\'e Paris Diderot, 75205 Paris Cedex 13, France} 
\author{Andrejs Cebers}
\affiliation{Institute of Physics, University of Latvia, Riga, Latvia}
\author{Jean-Claude Bacri}
\affiliation{Laboratoire Mati\`ere et Syst\`emes Complexes, UMR 7057\\ CNRS $\&$ Universit\'e Paris Diderot, 75205 Paris Cedex 13, France} 
\author{Jean-Marc Di Meglio}
\affiliation{Laboratoire Mati\`ere et Syst\`emes Complexes, UMR 7057\\ CNRS $\&$ Universit\'e Paris Diderot, 75205 Paris Cedex 13, France} 
\author{Charlotte Py}
\altaffiliation[Corresponding author ]{}
\email{charlotte.py@univ-paris-diderot.fr}
\affiliation{Laboratoire Mati\`ere et Syst\`emes Complexes, UMR 7057\\ CNRS $\&$ Universit\'e Paris Diderot, 75205 Paris Cedex 13, France}

\date{\today}

\begin{abstract}
We present an artificial swimmer consisting in a long cylinder of ferrogel which is polarized 
transversely and in opposite directions at each extremity. When it is placed
on a water film and submitted to a transverse %
oscillating magnetic field, this artificial {\it worm} undulates and swims.
Whereas symmetry breaking is due to the field gradient, the undulations of the worm result from
a torsional buckling instability as the polarized ends tend to align with the applied magnetic field. The
critical magnetic field above which buckling and subsequent swimming is observed may be predicted using
elasticity equations including the effect of a magnetic torque. As the length of the worm is varied, several undulation modes are observed which are in good agreement with the
bending modes of an elastic rod with free ends.
\end{abstract}

%\keywords{}

\maketitle

The design of robots that mimic animal locomotion has gained in interest during the last two decades, both for their medical applications  as well as for exploration tasks or search and rescue missions.  Among the various locomotion strategies of animals, the undulation locomotion is of particular interest for its simplicity and robustness, even on rough terrain.  The locomotion of snakes, eels or worms have been particularly studied from fundamental as well as biomimetic points of views \cite{Hu,Gillis,Korta} and robots inspired by their displacement methods have been proposed \cite{Abbott,Crespi}. To wirelessly power and control such robots, smart materials have been developed, often actuated by magnetic fields. The undulation of artificial swimmers are generally obtained by oscillating a magnetic head attached to a flexible tail \cite{Dreyfus,Sudo,Guo}. Beyond the technological challenge, such swimmers also raise a more fundamental interest on the description of the non-reversible flow around the body \cite{Purcell,Avron,Hosoi,Becker}. In this paper, we present a new and non-trivial undulation mechanism that is based on the torsional buckling instability of a worm-like swimmer under an oscillating magnetic field.

%Unlike macroscale swimmers, microorganisms, such as bacteria, eukaryotic cells or nematode worms, do not make use of  inertia to propel themselves by pushing backward the surrounding liquid and have developed various alternative locomotion strategies \cite{Taylor,Lighthill, Berg, Alexander, Korta}. At low Reynolds number, viscous forces dominate over inertia and cyclic motions with a single degree of freedom cannot  lead to net displacements, as popularized by the so-called scallop theorem of Purcell [6].  To break the time symmetry, Purcell proposed the idea of two-hinged swimmers performing a precise sequence of deformation. This inspired a large number of studies involving the prediction of the displacement direction, the efficiency of swimming and the practical design of such swimmers \cite{Hosoi,Becker,Najafi,Avron05}. In nature, symmetry breaking is achieved through two major mechanisms, namely the propagation of a helical wave through the rotation of a flagellum \cite{Berg1973,Purcell97} or the elastic deformation of the body \cite{Pironneau,Wiggins}. These locomotion modes remain a highly active research area, both for their theoretical aspects \cite{Avron,Hosoi,Becker} and for the design of bio-mimetic microrobots \cite{Abbott,Dreyfus}.

\begin{figure}[ht]
    %\centering
        \includegraphics[width=0.40 \textwidth]{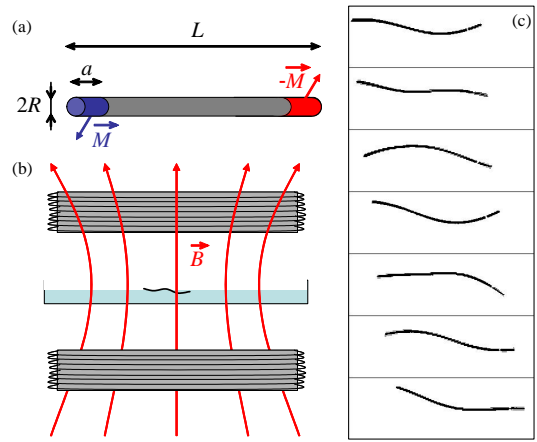}
    \caption{(a) Sketch of the magnetic worm-like swimmer: the colored zones represent the polarized regions. (b) Experimental set-up: a flexible magnetic worm is placed at the surface of water, between two coils in Helmholtz configuration, generating a vertical magnetic field $B$. (c) Top view of the swimmer under an oscillating magnetic field ($B=$~50~G, $f=$~1~Hz). The images are thresholded for better visualization. The time step between successive images is 0.33~s, and the length $L$ of the swimmer is 35 mm.}
    \label{figure1}
\end{figure}

%In this letter we present a novel swimmer, able to propel itself in an oscillating magnetic field, through a non-trivial and original mechanism. 
Our swimmer consists in a flexible cylinder of PVA (poly-(vinyl-alcohol), Sigma-Aldrich) based ferrogel, reticulated with glutaraldehyde (Sigma-Aldrich), in which $10\%$ in weight of ferromagnetic particles (black iron oxide), with micrometric size, are added.
 The particles are embedded in the gel, which has   a Young modulus $E= 1.36$~kPa (measured by indentation on a TA-XT2 texturometer). Before reticulation, the gel is poured into a cylindrical mold, with diameter $2R=1$ mm and length $10<L<50$ mm. Once the gelation is achieved, the gel is taken out of the mold and rinsed with water to remove free PVA chains. 
In order to provide permanent magnetic properties to the swimmer, a magnet is then placed  at each of its extremities, with opposite polarization directions. Over a distance $a=5$ mm from each extremity, the 
embedded particles acquire a permanent magnetization density $M$, oriented perpendicularly to the main axis of the cylinder, with opposite direction, see Fig.~\ref{figure1}(a). The magnetization density was measured on the magnetization curve according to Foner's method \cite{Foner} and equals 1.5~G.
The magnetic worm-like swimmer is then deposited at the surface of water in a Petri dish, set between two coils in Helmholtz configuration (Fig.\ref{figure1}(b)). When we supplied with AC, the coils generate a vertical oscillating magnetic field, with adjustable magnitude $B$ and frequency $f$.

\begin{figure}[ht]
    %\centering
        \includegraphics[width=8cm]{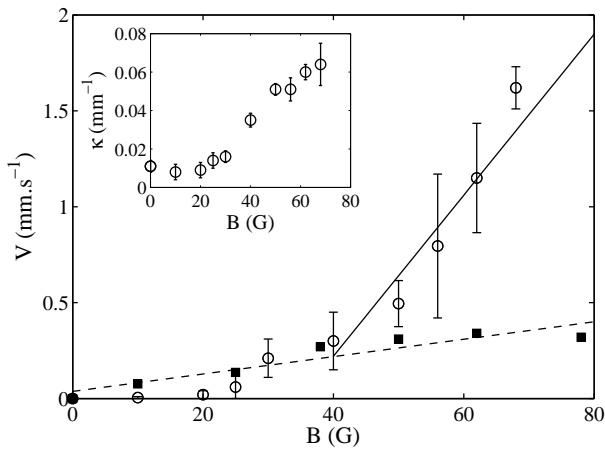}
    \caption{Swimming velocity $V$ of the worm as a function of the magnitude $B$ of the applied magnetic field,  for a constant field (squares) and for an oscillating field at $f=1$ Hz (circles). The length of the swimmer is $L=15.6$ mm. The critical field, $B_c$, is  here 40~G. Inset: maximum curvature $\kappa$ of the swimmer as a function of the magnetic field, for $f=1$ Hz. }
    \label{figure2}
\end{figure}

Submitted to a constant magnetic field ({\it i.e.} for $f=$ 0 Hz), the worm drifts slowly from the center to the edge of the  Petri dish, without deformation \cite{EPAPS}. The displacement velocity $V$ is very small ($\approx$ 0.2 mm.s$^{-1}$) and proportional to the magnitude of the magnetic field $B$, Fig. \ref{figure2}. The motion is  due to the existence of a gradient of the horizontal component of the magnetic field in the radial direction, which increases by 30 $\%$ from the center  to the edge of the dish. A magnetic force, proportional to the field gradient, and thus to the field magnitude, acts on all the ferromagnetic particles of the worm (not only its extremities) and pushes it away from the center of the Petri dish.

Under an oscillating magnetic field, the worm undulates (Fig.~\ref{figure1}(c)) and swims toward the edge of the dish with a much higher velocity, see  \cite{EPAPS}. This only occurs above a critical value of the applied magnetic field, $B_c$ (Fig.~\ref{figure2}).  Below this value, the swimmer drifts slowly without undulating, just as when subjected to a constant magnetic field.

In order to investigate the origin of the undulation, white dots were drawn along the worm body to serve as labels. The dots allow to highlight the full 3D deformation of the swimmer and reveal that the worm actually twists around its main axis when it undulates \cite{EPAPS}. The magnetic moments of the extremities tend to align with the applied vertical magnetic field, and since they are polarized in opposite direction, this results in the twist of the ferrogel cylinder. Above a certain threshold, twisting of an elastic rod is known to lead to its buckling through an elastic instability, this is known as the Greenhill problem \cite{greenhill}: a twisted elastic rod buckles into a 3D helix. As the magnetic swimmer is here constrained in the plane of the free surface of the liquid by surface tension, the deformed shape of the worm is a confined helix, a quasi-sinusoid \cite{EPAPS}.  The alternative twisting of the worm under the oscillating magnetic field would therefore be at the origin of the bending wave.

\begin{figure}[ht]
    %\centering
        \includegraphics[width=8cm]{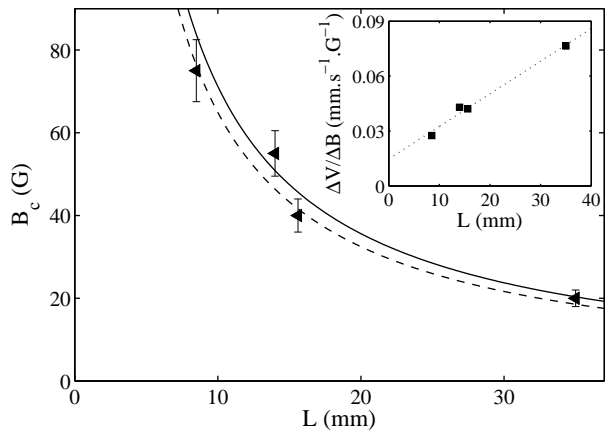}
    \caption{Critical field $B_{c}$ as a function of the length $L$ of the swimmer ($f=~1~$Hz). Dash line: fit of the data, of the form $B_c=\alpha/L$ ($\alpha=649$ G.mm); solid line: model, Eq. \ref{eq_Hc}, using the experimental values of parameters (no adjustable parameter): $E=$~1.36~kPa, $R=$~0.5~mm, $a=$~5~mm, $M=$~1.5~G. Insert: variation of velocity beyond the threshold with the applied magnetic field as a function of the length of the swimmer (dotted line: linear fit).}
    \label{fig_Hc}
\end{figure}

To test this hypothesis, the critical field $B_c$ was measured for worms of different lengths (Fig.~\ref{fig_Hc}). For an elastic rod with Young modulus $E$ and twist modulus $C$, the total elastic energy reads:
\begin{equation}
E_{el}=\frac{1}{2}EI\int_0^{L}\, \left( \frac{d\theta}{ds} \right)^{2}ds+\frac{1}{2} C \int_0^{L}\,\left( \frac{d\phi}{ds} \right)^{2}ds, \label{eq_elasticNRJ}
\end{equation}
with  $I=\pi R^4/4$  the area moment of inertia and $s$ a curvilinear abscissa along the rod \cite{Landau}.
The first term corresponds to the bending energy, $\theta$ being the local bending angle, and the second term is the torsion energy, with $\phi$ the local twisting angle. Both terms are proportional to $EI/L$. The rod is submitted to a twisting torque due to the interaction of the magnetic moments of the extremities with the applied magnetic field, whose modulus reads
$T= \| \pi R^2 a \overrightarrow{M} 
\wedge \overrightarrow{B} \|$, 
with $\pi R^2 a$ the volume of the polarized region, and $M$ the magnetization density.  Comparing the elastic and magnetic energies shows that the critical twisting torque above which buckling occurs should scale as $EI/L$ and thus the critical magnetic field $B_c$ as $E R^2 /( a M L)$. A more detailed description of the problem, based on the elasticity equations coupled  with a magnetic torque, allows to obtain the missing pre-factor in the above expression. The critical field is then found as \cite{EPAPS}:
\begin{equation}
B_c=   \frac{\pi E R^2}{ 2 a M L}. \label{eq_Hc}
\end{equation}

This model is compared to the experimental data of $B_c$ as a function of the swimmer's length, see Fig.~\ref{fig_Hc}. A good qualitative and quantitative agreement is found, supporting that the twisting of the worm due to the interaction of its polarized extremities with the applied magnetic field is indeed at the origin of the undulation of the swimmer. 
%a mettre ici, ou ds la conclusion?
The scaling of the swimming threshold, $B_c \sim R^2/ aL$, is favorable to miniaturization since it is globally independent on the swimmer's dimensions.
The propagative nature of this bending wave, allowing the worm to swim, originates from the asymmetry of the twisting torque on both extremities, resulting from the  radial gradient of the applied field. This gradient is here inherent to the experimental set-up. Another way of breaking the head/tail symmetry is to unbalance the magnetizations of the extremities (see \cite{EPAPS}).

Above the critical field $B_c$, the swimming velocity of the worm increases with the applied field, with a slope twenty times larger than below $B_c$ (Fig.~\ref{figure2}). This increase is associated with a rise in the body curvature  with $B$, Fig.~\ref{figure2} (insert). 
%Above $B_c$, the rise in the velocity with $B$ is indeed due to an increase in the applied twisting torque and thus in the bending of the worm through the elastic instability. 
As the total elastic energy ($\sim EI/L$) is inversely proportional to its length, this effect is all the more pronounced as the length of the swimmer is larger (Fig.~\ref{fig_Hc} insert). The increase in the slope $\Delta V / \Delta H$ with $L$ is in fact due to the sum of two effects: the increase in the flexibility of the rod with $L$, and the fact that the gradient of the magnetic field and thus the difference of magnetic torque applied at both extremities (at the origin of the symmetry breaking) also increase with the length of the swimmer.

\begin{figure}[ht]
   \centering
        \includegraphics[width=8cm]{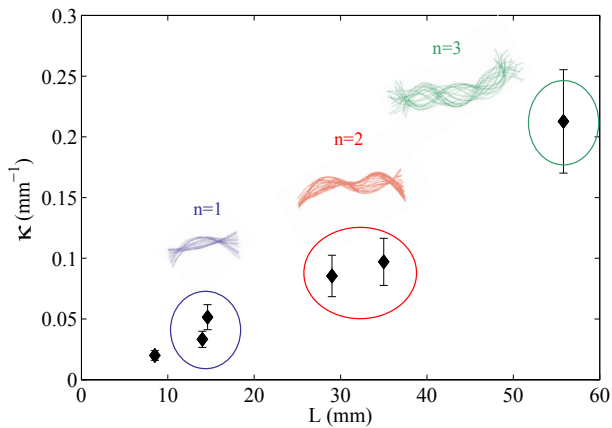}
    \caption{Curvature of the worm as a function of its length, for $B=$ 50 G and $f=$ 1 Hz. The inserts are the superpositions of the successive shapes of the worm over one period, revealing the successively obtained mode shapes.}
   \label{fig_modes}
\end{figure}

Interestingly, as the length of the swimmer is increased (for a fixed value of $B$ and $f$), different undulation modes are observed and characterized by $n$,  the number of antinodes, Fig.~\ref{fig_modes}. These successive modes of deformation are associated with an increasing curvature for increasing $L$. Note that modes 2 and 3 cannot be explained by a static twisting-buckling instability as they would require a twisting angle larger than $\pi$ which, in practice, is impossible here: if the magnetic moments of the extremities get aligned with the applied field (which happens for a torsion angle of $\pi$), the magnetic torque cancels and no further rotation can happen. Another type of explanation has therefore to be sought through the dynamic behavior of the swimmer.

\begin{figure}[ht]
    %\centering
        \includegraphics[height=4.2cm]{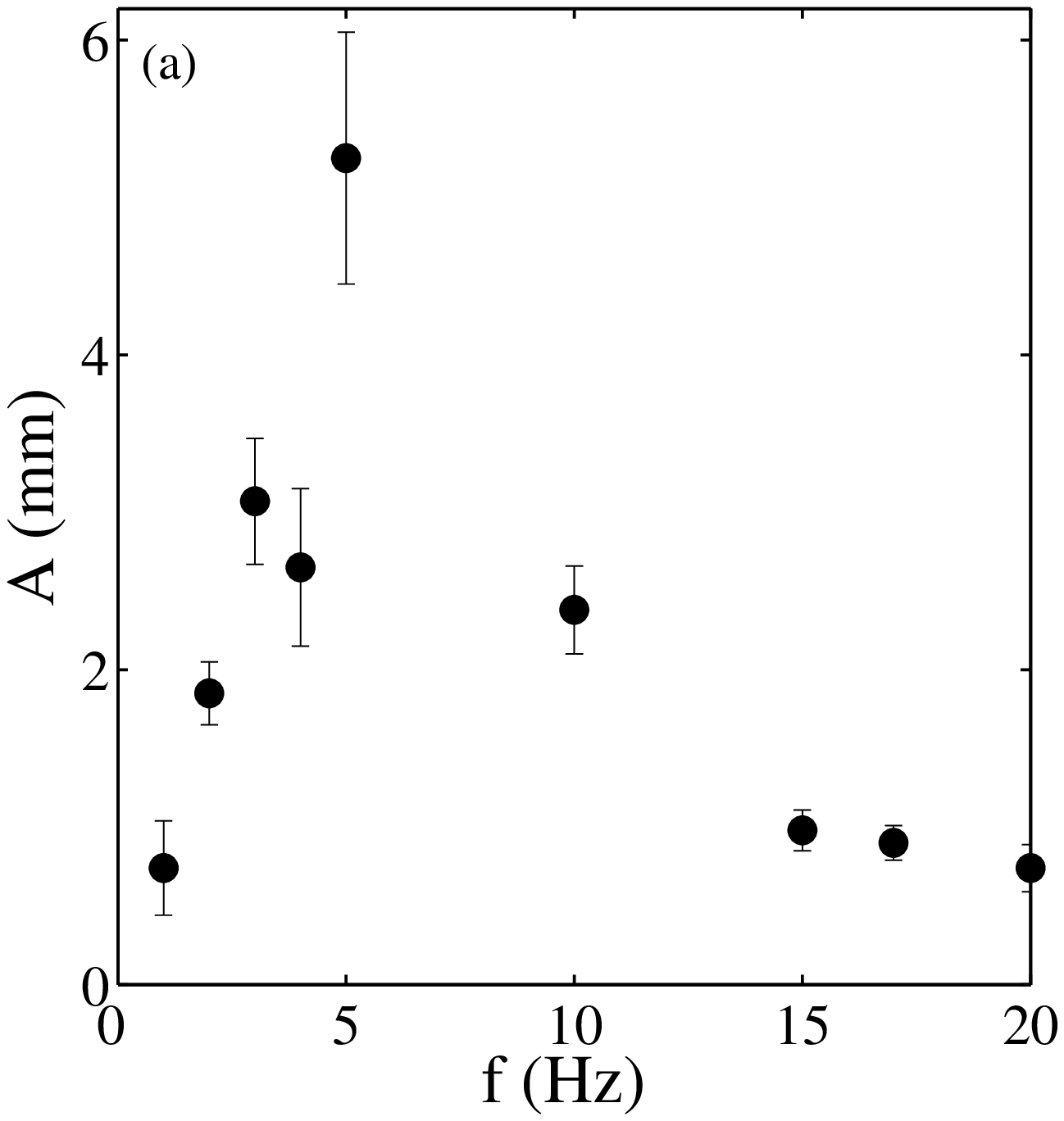}
\hspace{0.01cm}
        \includegraphics[height=4.2cm]{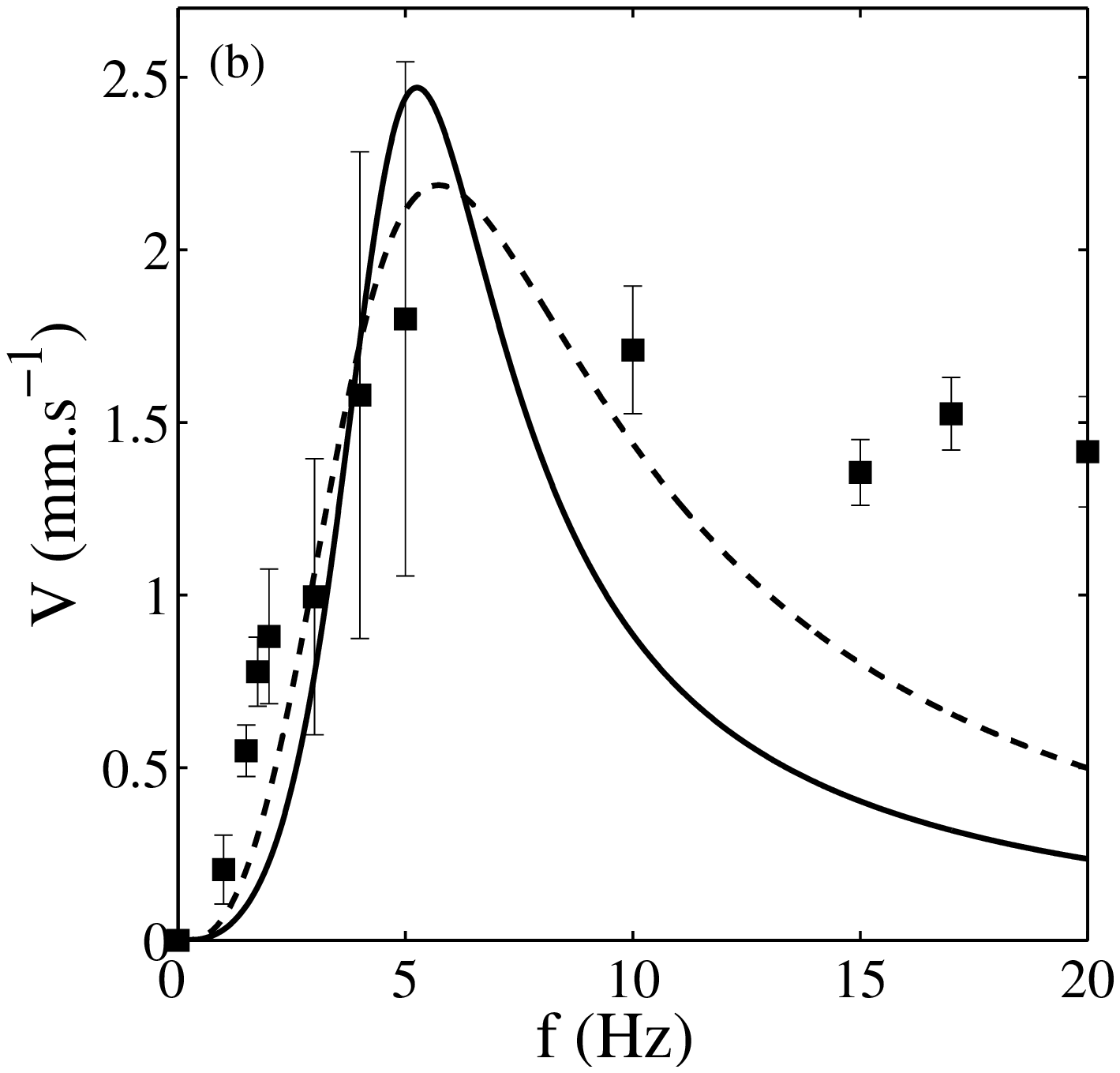}
    \caption{Bending amplitude $A$ (a) and velocity $V$ (b) of the swimmer (the $f=0$~Hz drift velocity, 0.23~mm.s$^{-1}$, has been substracted) as a function of the frequency $f$ of the oscillating magnetic field ($B=$ 50 G, $L=$ 29 mm).  Solid line: model (Eq. \ref{eq_Alexander})  with the sole mode 2; dash line:  model with  mode 2 up to 10 Hz and  mode 3 above 15Hz.}
    \label{fig_A_V}
\end{figure}

The bending amplitude $A$ and swimming velocity $V$ are measured as a function of the frequency of the applied magnetic field, for fixed values of $L$ and $B$,  Fig.~\ref{fig_A_V}. Both $A$ and  $V$ increase approximately linearly with $f$ up to 5~Hz, where they reach a maximum, and then decrease. The curve of the amplitude displays a pronounced peak which resembles a resonance curve. The peak frequency  may be compared with the free vibration frequencies of an elastic rod. Two main modes of vibrations could be relevant here: bending modes and torsion modes (or coupling between both). 
For bending modes, the boundary conditions for a rod with free extremities in mode 2 (the  mode observed at all frequencies in Fig.~\ref{fig_A_V}) impose $kL=7.85$, with $k$ the wavenumber. Using the dispersion relation $\omega = \beta k^2$ ($\beta= \sqrt{EI/(\rho \pi R^2)}$, with  $\rho=1050$ kg.m$^{-3}$ the mass per unit volume of the rod), the eigenfrequency is found as \cite{Lalanne}
\begin{equation}
f_{n=2} = \frac{9.82 \beta}{L^2}. \label{eq_f2}
\end{equation}

 Using the previously mentioned values of the parameters (and $L=29$ mm), $f_{n=2}$ is here equal to 3.44~Hz, close to the experimental peak frequency at 5~Hz. The eigenfrequency of a torsion rod, for $n=2$, is of the order of $230$~Hz:  the torsion modes alone are therefore not relevant to account for the dynamics of the magnetic swimmer. However, coupling between the bending modes (which are predominant) and the torsional modes is needed to explain that the peak frequency is higher than the eigenfrequency of  pure bending.

The evolution of the swimming velocity with the applied frequency may now be explained thanks to a classical kinematic model. The cutting frequency between inertial and viscous regime, $f_c=2\eta /\rho \pi r^2$  (with $\eta=10^{-3}$ Pa.s$^{-1}$ the viscosity of water),  is here of the order of 2.6 Hz: over the frequency range studied here, we are therefore in a mixed regime where viscous and inertial effects coexist. As the progression of the worm happens at relatively low Reynolds numbers (from 0.3 to 2  and from 1 to 20, respectively for the longitudinal and transverse motion), we use the predictions of the swimming velocity in the $Re \rightarrow 0$ limit although the bending motion of the swimmer is mainly dominated by inertia.
Balancing the longitudinal forces acting on a sinusoidal elongated worm, submitted to purely viscous friction, allows to express the progression velocity as \cite{Alexander, note,Gray-Hancock}
\begin{equation}
V= \frac{\omega}{k} \frac{\tilde{A}^2k^2 (c -1)}{\left( 2 + \tilde{A}^2k^2 (c-1/2) \right)}
\label{eq_Alexander} ,
\end{equation}
with $\tilde{A}$, $\omega=2 \pi f$, and $k$, respectively the amplitude, pulsation and wavenumber of the propagative wave, and $c$ the ratio of the transverse to longitudinal friction coefficients ($c=2$ \cite{Lighthill,Lauga}). In the above equation, $k$ is obtained from the wavelength measured on the superposition of the successive shapes of the swimmer (as shown Fig. \ref{fig_modes}). 
The main mode of deformation is mode 2 for all frequencies up to 10 Hz, but a combination of modes 2 and 3 is observed above 15 Hz, leading to two possible values of the wavenumber in Eq. \ref{eq_Alexander} at high frequency.
 The amplitude $\tilde{A}$ to be considered in Eq. \ref{eq_Alexander} is that of the propagative component of the bending wave animating the worm: the wave is indeed mainly stationary in our case, with a propagative part  estimated here to be $18\%$ of the total measured amplitude (based on a gradient of magnetic field of $18\%$ along the swimmer's body). 
This model is compared to the experimental results on Fig.~\ref{fig_A_V}(b) and shows a reasonable qualitative agreement,  given the fact that there are no adjustable parameters. The main features of the swimming dynamics can therefore be explained by a simple kinematic model based on viscous friction associated with the bending modes of an elastic rod. Considering the mode 3 above 15~Hz (dash line) gives a better agreement than considering the mode 2 for all frequencies (continuous line); however, the model under-estimates the swimming velocity at high frequency.  An alternative model has been developed, where viscous friction is considered for the longitudinal motion and inertial friction for the transverse motion, without great influence on the quantitative results.

In conclusion, we have designed a novel magnetic swimmer that moves through an original mechanism: generation of a bending wave via a twisting/buckling instability, and propagation of this wave by breaking the head-tail symmetry of the applied twisting torque. The swimming dynamics is then mainly governed by the bending modes of the equivalent elastic rod.
%Although the experiments are carried on at the macroscopic scale, the swimming mechanism is fully adaptable at the microscopic scale.

\textbf{\normalsize{Acknowledgements }} \\
This work was supported by the Agence Nationale de la Recherche (grant {\it Locomoelegans} $NT05-3-41923$). The authors thank Howard Stone for helpful discussion.


\begin{thebibliography}{25}
\bibitem{note}
Note that the integration is here performed at second order at all steps, leading to a slightly different expression than the one in \cite{Alexander}.
\end{thebibliography}
\end{document}